\documentclass[a4paper]{article}
\usepackage[letterpaper, portrait, margin=2cm]{geometry}
\usepackage[english]{babel}
\usepackage[utf8]{inputenc}
\usepackage{amsmath}
\usepackage{mathtools}
\usepackage{graphicx}
\usepackage{upgreek}
\usepackage[colorinlistoftodos]{todonotes}
\DeclareMathAlphabet{\mathcal}{OMS}{cmsy}{m}{n}
\SetMathAlphabet{\mathcal}{bold}{OMS}{cmsy}{b}{n}
\usepackage{braket}
\usepackage{url}
\usepackage{float}
\usepackage{array}
\usepackage{authblk}
\usepackage{graphicx, caption}
\usepackage{amsfonts}
\usepackage{braket}
\usepackage{amsmath,bm}
\usepackage{svg}
\usepackage[super,comma,square]{natbib}
\usepackage{graphicx}
\usepackage{hyperref}

\usepackage{amssymb}
\usepackage{amsthm}
\usepackage{authblk}
\usepackage{color}
\usepackage{mathtools}

\usepackage{bbm}
\usepackage{tikz}
\usepackage{setspace}

\usepackage{abstract}

\usepackage{qcircuit}

\usepackage{blindtext}
\usepackage{multicol}
\usepackage[font=small,labelfont=bf]{caption}

\begin{document}

\title{The Impact of Hardware Specifications on Reaching Quantum Advantage in the Fault Tolerant Regime}
\author[1,2]{Mark Webber}
\author[3]{Vincent Elfving}

\author[1,2]{Sebastian Weidt}
\author[1,2]{Winfried K. Hensinger}
\affil[1]{\small{\textit{Sussex Centre for Quantum Technologies, Department of Physics and Astronomy, University of Sussex,
Brighton, BN1 9QH, United Kingdom}}}
\affil[2]{\small{\textit{Universal Quantum, Brighton, United Kingdom}}}
\affil[3]{\small{\textit{Qu $\&$ Co B.V., Amsterdam, The Netherlands}}}

\date{\today}

\maketitle

\renewcommand{\abstractname}{}    % clear the title
\renewcommand{\absnamepos}{empty} % originally center

\begin{abstract} 

We investigate how hardware specifications can impact the final run time and the required number of physical qubits to achieve a quantum advantage in the fault tolerant regime. Within a particular time frame, both the code cycle time and the number of achievable physical qubits may vary by orders of magnitude between different quantum hardware designs. We start with logical resource requirements corresponding to a quantum advantage for a particular chemistry application, simulating the FeMoco molecule, and explore to what extent slower code cycle times can be mitigated by using additional qubits. We show that in certain situations architectures with considerably slower code cycle times will still be able to reach desirable run times, provided enough physical qubits are available. We utilize various space and time optimization strategies that have been previously considered within the field of error-correcting surface codes. In particular, we compare two distinct methods of parallelization, Game of Surface Code's Units, and AutoCCZ factories, both of which enable one to incrementally speed up the computation until the reaction limited rate is reached. Finally we calculate the number of physical qubits which would be required to break the 256 bit elliptic curve encryption of keys in the Bitcoin network, within the small available time frame in which it would actually pose a threat to do so. It would require approximately 317 million physical qubits to break the encryption within one hour using the surface code, a code cycle time of 1 $ \mu s$, a reaction time of 10 $ \mu s$, and physical gate error of $10^{-3}$. To break the encryption instead within one day it would require 13 million physical qubits.

% We investigate the impact of the average number of T gates per layer  and identify the optimal value for particular scenarios.  If the base physical error rate was instead the more optimistic value of $10^{-4}$, 33 million physical qubits would be required to break the encryption in 1 hour. 

\end{abstract}

\section{Introduction}

%tim eoptimal limit reference fowler - https://arxiv.org/pdf/1210.4626.pdf "T depth of a quantum circuit as the minimum number of T gates that need to be implemented sequentially to complete execution
% reference ZXXZ???
%https://arxiv.org/pdf/1905.08916.pdf

%GoSC- The number of rotations is the T count and the number of layers is the T depth. Each rotation can be performed by consuming a magic state via a Pauli product measurement.

With the advent of quantum computers, the race to a quantum computational advantage has gained serious traction in both the academic and commercial sectors. Recently, quantum supremacy has been claimed \cite{Arute2019, Deshpande2021QuantumSampling} on quantum devices with tens of qubits. However, the targeted problems solved were theoretical in nature, and not relevant industrial applications. Quantum advantage, conversely, is a stronger form of supremacy that shows an industrially relevant computational problem solved in a reasonable time-frame that would be practically impossible to do using any classical supercomputer. There is a large physical qubit overhead associated with quantum error correction, which is required to run some of the most powerful algorithms.  There are many factors which will determine the ability for different quantum computing architectures to scale.  Consequently, within a given time frame the maximum size (qubit count) of a device could vary by orders of magnitude. The code cycle time (base unit of operation in the surface code) will also vary considerably between platforms. Therefore it is of interest to investigate the interplay between the code cycle time and the number of achievable qubits, and the resulting impact on the feasibility for a particular device to achieve a quantum advantage. We calculate the physical qubit and run time requirement for relevant problems in chemistry and cryptography with a surface code error corrected quantum computer, comparing parameters typical to various types of hardware realizations. 

%We extend the scope further to general estimates in terms of the required logical qubits and non-transversal logical operations.  When layers of T gates are parallelized, the number of physical qubits within a particular device will play an important role in determining the feasibility of reaching desirable run times, in addition to the more usually considered code cycle time. 

%In particular, we focus our attention on the impact of the cost of scalability for achieving a quantum advantage using a fault-tolerant quantum computer, comparing parameters typical to various types of hardware realizations.%

Algorithms which are tailored to NISQ devices generally consist of a hybrid approach, where a low depth circuit is parameterized, and iterated through a classical optimizer. These NISQ algorithms are more often heuristic in nature than their fault tolerant counterparts and so providing rigorous resource estimates can be more challenging. Many of the most powerful quantum algorithms require a circuit depth which greatly exceeds the capabilities of NISQ era devices, and for some applications the number of required logical qubits and operations are known. The quantum threshold theorem states that a quantum computer using error correction schemes, and a physical error below a certain threshold, can suppress the logical error rate to arbitrarily low levels \cite{Aharonov2008Fault-TolerantRate,Knill1998ResilientThresholds,Kitaev2003Fault-tolerantAnyons}. Therefore one could run an algorithm with an arbitrarily long circuit depth provided enough qubits are available to perform the required level of error correction. There is a large time overhead for performing logical operations at the error corrected level relative to operations performed on physical qubits. For the foreseeable future, classical computers will have a clock rate that is orders of magnitude faster than error corrected quantum computers. To determine the problem size at which a quantum computer will outperform a classical computer one must consider both the algorithmic speed up as well as the relative difference between their associated clock rates. By making use of parallelization schemes, quantum computers can speed up the effective rate of logical operations at the cost of additional qubits, and so the ability to scale to large enough device sizes will also play a role in determining the feasibility of reaching desirable run times.

%We explore the interplay between the hardware's code cycle time, and it's scalability by utilizing some of the latest developments in the surface code, with a focus on parallelizing T gate layers of the algorithm at the cost of additional physical qubits.

The surface code \cite{Fowler2012SurfaceComputation,Bravyi1998QuantumBoundary,Dennis2002TopologicalMemory} is the most researched error correction technique, particularly in regards to end-to-end physical resource estimation. There are many other error correction techniques available and the best choice will likely depend on the underlying architecture's characteristics, such as the available physical qubit-qubit connectivity. Superconducting qubits are one of the leading quantum computing platforms and these devices generally consist of static qubits arranged on a grid where only nearest neighbour interactions are natively available. Long distance connections must be enabled by sequences of nearest neighbour swap operations, which in the context of a NISQ device may limit their computational power \cite{Cross2019, Webber2020EfficientComputer}. The limited connectivity of superconducting qubits in part motivated the continued research into the surface code which relies only on nearest neighbour interactions between physical qubits arranged on a 2D grid. In the following we briefly introduce some of the alternative error correction techniques to the 2D surface code. Error correction codes which rely on global interactions at the physical level have favourable encoding rates as a function of code distance \cite{roffe2020decoding} but enabling this global connectivity on large devices may be challenging, or the connectivity overheads may outweigh the benefits relative to closer-range-connectivity codes. Entanglement distribution may be a viable method of enabling distant connectivity for large scale devices with limited physical connectivity in the limit of large reserves of quantum memory \cite{deBeaudrap2019QuantumArchitectures}. Higher dimensional error correction codes can have access to a greater range of transversal gates \cite{Campbell2017, VasmerThree-dimensionalArchitectures} which may considerably improve final run-times, where transversal implies that each qubit in a code block is acted on by at most a single physical gate and each code block is corrected independently when an error occurs. Realising this 3D (or greater) physical connectivity could be challenging for many of the current quantum platforms; photonic-interconnected modules may be the most flexible architecture with regards to its possible connectivity graph \cite{Monroe2012}, however, currently achieved connection speeds would present a considerable bottleneck \cite{Stephenson2020High-RateNetwork}.  A variant of the 3D surface code may still be realisable with hardware that is only scalable in two dimensions because the thickness (extra dimension) can be made relatively small and independent of code distance \cite{ScrubyNumericalCode}. In section \ref{s2} we highlight the surface code in more detail and include relevant considerations for physical resource estimation. 

%In this manuscript we investigate the impact of hardware considerations such as code cycle time and the cost of scalability and their interplay which results from surface code strategies that allow one to linearly trade physical qubits for run-time by parallelizing T gate layers. 

Here we highlight some of the leading quantum computing platforms, their relevant error correction strategies, and contrast their rate of operations. The surface code is the front-running proposal for error correction in superconducting devices and their associated code cycle times (the base sequence of hardware operations) have been estimated to be in the range of 0.2 $ \mu s$-10 $ \mu s$ \cite{Fowler2012, Litinski2019gameofsurfacecodes}. There is a great variety within different implementations of trapped ion architectures, particularly with regards to the method of enabling connectivity. A proposed scalable trapped ion design that relies on shuttling to enable connectivity, and microwave based gates has estimated the code cycle time to be 235 $ \mu s$ \cite{Lekitsch2017}. For this shuttling based design alternative error correction protocols may be more effective than the surface code due to the variable-mid range connectivity that is possible, but in this work we constrain ourselves to the surface code. Small trapped ion modules connected via photonic interconnects have been envisaged with the surface code \cite{Nigmatullin2016MinimallyComputing}, but due to their extremely flexible connectivity, higher dimensional error correction codes may one day be utilized. The code cycle time for trapped ions with photonic interconnects would depend on how the physical qubits are distributed across modules; one approach advocates for two tiers of encoding to minimize the use of the slower interconnects \cite{Li2016HierarchicalSize}. Trapped ion architectures have achieved some of the highest gate fidelities to date \cite{Srinivas2021High-fidelityQubits, Schafer2018, Ballance2016, Gaebler2016High-FidelityQubits, Webb2018}; the lower the base physical error rate, the lower the code distance will need to be in the error correction protocol and therefore fewer physical qubits would be needed per logical qubit.  A fault tolerant Silicon based architecture has been proposed using the surface code with code cycles estimated to be 1.2 $ms$ \cite{Ogorman2016AComputer}. The error correction choice for photonic devices will depend on the underlying design; the primary candidate for a particular fault tolerant proposal \cite{BourassaBlueprintComputer} is the RHG lattice \cite{Raussendorf2004Long-rangeStates, Raussendorf2006AComputer}. The degree to which an architecture is scalable will vary greatly between architecture types and within particular stages of development. 

%Modularity is a key component of scalability; a successful design could rapidly increase the number of physical qubits once the base building block is established and then iterated.

In this work we provide physical resource estimates to achieve a quantum advantage with a quantum computer using the surface code. Using some of the latest time optimization strategies \cite{Litinski2019gameofsurfacecodes, Gidney2019HowQubits, Gidney2019FlexibleStates}, we investigate the interplay between the code cycle time and the number of achievable physical qubits in the underlying architecture. In the following sections, we focus on quantum algorithms which have been hypothesized to offer a disruptive quantum advantage for industry-relevant applications. We first provide a brief overview of quantum computing for chemistry, and highlight the logical resource requirements for a quantum advantage use case, simulating the ground state of the FeMoco molecule, which we use as the starting point for our investigation. We then calculate the number of physical qubits that are required to break the elliptic curve encryption of Bitcoin keys within the time frame that it actually poses a threat to do, as a function of the code cycle time.

\subsection{Fault tolerant quantum chemistry}
When numerically simulating quantum chemistry problems, typically a set of independent functions, known as a basis set, are introduced to describe the physical wave function of the system. This introduction does not remedy the exponential increase of parameters with system size but enables one to balance the achievable accuracy against the required computational resources. Richard Feynman was perhaps the first to envisage a quantum computer and its application to the simulation of physics and chemistry \cite{Feynman1982SimulatingComputers}. It is now expected that quantum computers will eventually be able to perform electronic structure calculations with a quality of solution typical to the most accurate classical methods but with run times comparable to the approximate techniques, such as density functional theory.

The Quantum Phase Estimation (QPE) algorithm generates eigenvalues for a general unitary operator and it can be applied to quantum chemistry to find the eigenenergies of chemistry Hamiltonians to FCI (full configuration interaction, i.e., exact) precision. Unlike the Variational Quantum Eigensolver (VQE) \cite{Peruzzo2013} which involves many iterations ($\mathcal{O}(1 / \epsilon^2)$ with accuracy $\epsilon$)  of low depth circuits, the QPE algorithm requires $\mathcal{O}(1)$ iterations of a circuit with a depth scaling as $\mathcal{O}(1 / \epsilon)$. The large depth required in the QPE algorithm means that it will only be possible with error corrected devices, because NISQ devices would lose their coherence long before the end of the circuit. 

Hamiltonian simulation is used as a subroutine in the quantum phase estimation algorithm, and it involves constructing a quantum circuit which approximates the evolution of the input state according to the Hamiltonian. Two of the main paradigms for Hamiltonian simulation are Trotterization and Qubitization.  Qubitization \cite{Low2019HamiltonianQubitization, Berry2019QubitizationFactorization} can be used to simulate the Hamiltonian evolution by using quantum signal processing \cite{Low2016OptimalProcessing} but more commonly it is used to generate a quantum walk \cite{Poulin2018QuantumCount} upon which one can directly perform phase estimation. Qubitization is perhaps the most favored method for simulating chemistry Hamiltonian dynamics because it achieves the provably optimal scaling in query complexity and approximation error albeit while requiring more logical qubits than other methods. 

%In this work we begin with the logical resource estimates provided by Elfving et al \cite{Elfving2020HowChemistry} for the chromium dimer at a CAS size of 26 at chemical accuracy.  Chemical accuracy is a commonly used standard of accuracy within the quantum chemistry field and it means that the error is no greater than 1 kcal/mol relative to a hypothetical exact energy or an experimental measurement. %This concept is distinct to that of precision which in this context is instead the computational error relative to a computational reference, for example, one can achieve chemical precision within a minimal basis set for a particular molecule, but even the exact answer within the minimal basis set would not necessarily be sufficient to achieve chemical accuracy.

Previous work has investigated the potential for quantum computers to provide a quantum advantage by performing ground state energy estimations on the catalytic complex known as FeMo-co \cite{Reiher2017, Li2018TheSimulations, vonBurg2020QuantumCatalysis,Lee2020EvenHypercontraction}. FeMo-Co is a large molecule expressed in biology which reduces  $N_2$ from the atmosphere, and a better understanding of this process could provide a significant commercial advantage by improving the efficiency of nitrogen fixation for the production of ammonia for fertilizer. In this work we start our investigation with the latest logical resource requirements for simulating FeMo-co and calculate the number of physical qubits required to reach a desirable run time as a function of the code cycle time of the hardware.

\subsection{Breaking Bitcoin's encryption}\label{1.2}
Bitcoin, the first decentralized cryptocurrency is continuing to grow in popularity. Bitcoin has properties which make it desirable as a hedge against inflation, for example, the rate of supply is known, decreases with time, and is entirely independent of demand. The decentralized nature of the blockchain makes it censor resistant and it can operate in a trustless manner. There are two main ways in which a quantum computer may pose a threat to the Bitcoin network \cite{AggarwalQuantumThem, TesslerBitcoinComputing}. The first and least likely is the threat to the proof of work mechanism (mining) for which a quantum computer may achieve a quadratic speedup on the hashing of the SHA256 protocol with Grover's algorithm \cite{Grover1996ASearch}. The algorithmic speedup is unlikely to make up for the considerably slower clock cycle times relative to state of the art classical computing for the foreseeable future \cite{TesslerBitcoinComputing}. The second and more serious threat would be an attack on the elliptic curve encryption of signatures. Bitcoin uses the Elliptic Curve Digital Signature Algorithm (ECDSA) which relies on the hardness of the Elliptic Curve Discrete Log Problem (ECDLP) and a modified version of Shor's algorithm \cite{Shor,HanerImprovedLogarithms, RoettelerQuantumLogarithms} can provide an exponential speedup using a quantum computer for solving this problem. Bitcoin uses ECDSA to convert between the public and private keys which are used when performing transactions. With best practices (using new addresses for each transaction), the only point at which a public key is available and relevant to a eavesdropper, is after a transaction has been broadcast to the network but prior to its acceptance within the blockchain. In this window, transactions wait in the ``mem pool" for an amount of time dependent on the fee paid, the time taken for this process is on average 10 minutes but it can often take much longer. Gidney and Eker{\aa} estimated that it would require 20 million noisy qubits and 8 hours to break the 2048 RSA encryption \cite{Gidney2019HowQubits} which is of a comparable difficulty to the EC encryption of Bitcoin. The maximum acceptable run time for breaking Bitcoin's encryption makes it particularly well suited to our investigation into the cost of parallelization and the interplay between the code cycle time and scalability, which we present in section \ref{s3}.

In the following section we introduce considerations for error correction and provide an overview of the space and time optimizations within the surface code that we make use of in this work.

\section{Space and time optimizations in the surface code}\label{s2}

%\begin{figure}[t]
%\centering
%\includegraphics[width=17.8cm]{doubled_1.png}
%\caption[Tof gate cost]{\label{fig:1} (A) The required time (in %days) for a ground state energy calculation, for $Cr_2$ at a CAS size of (26,26) to chemical accuracy with the sparse Qubitization algorithm, all as a function of inverse physical gate error with a code cycle time of 1$ \mu s$. We include the required time to solve this problem on a top 5 HPC with $\sim$125PFLOPS (estimated by extrapolating from a 1.2TFLOP (desktop PC) simulation at orbital numbers equal to 10, 12 and 14). (B) The required number of physical qubits for a ground state energy calculation, for $Cr_2$ at a CAS size of (26,26) to chemical accuracy with the sparse Qubitization algorithm, all as a function of inverse physical gate error.  }
%\end{figure}

In this section we briefly introduce quantum error correction in the context of resource estimation and explain some of the available strategies within a surface code setup which are selected based upon a preference for space (physical qubit requirement) or time (final run time of the algorithm). 

\subsection{The available gate set}
An important consideration for quantum error correction is the available logical gate set which is generally more restricted than the underlying physical gate set. The Clifford gates are those that map Pauli operators onto other Pauli operators, and the set can be generated by various combinations of the set $\{H,CNOT,S\}$ where the $S$ gate is the Pauli $Z^{1/2}$. The Gottesman-Knill theorem \cite{Gottesman1998TheComputers} states that any Clifford circuit of finite size can be simulated in polynomial time (efficiently) with a classical computer. The Clifford gate set in combination with any non-Clifford gate is sufficient for universal quantum computation, and two of the most commonly considered non-Clifford gates are the $T$ gate ($Z^{1/4}$) and the Toffoli gate (control-control-not). Any arbitrary angle single qubit gate can be decomposed into long sequences of the fixed angle $H$ and $T$ gates with chain length scaling with desired precision, as per the Solovay-Kitaev theorem \cite{Dawson2006TheAlgorithm}.

\begin{equation}
T = \begin{pmatrix}
1 & 0 \\
0 & e^{i\pi/4} 
\end{pmatrix},\:\:
S = \begin{pmatrix}
1 & 0 \\
0 & i 
\end{pmatrix},\:\:
    H = \frac{1}{\sqrt{2}}\begin{pmatrix}
1 & 1 \\
1 & -1 
\end{pmatrix},\:\:
CNOT= \begin{pmatrix}
1&0&0&0 \\
0&1&0&0 \\
0&0&0&1 \\
0&0&1&0 \\
\end{pmatrix}.
\end{equation}

 The surface code has transversal access to the CNOT, and the H and S gates can be realized with a low overhead using other techniques \cite{Litinski2018LatticeCodes}. The T-gate is not transversal in the surface code and it must be effectuated using methods which incur a large space-time volume overhead relative to the other gates listed here. The T gate can be constructed by consuming a magic state, $\ket{m} = (\ket{0} + e^{i\pi/4}\ket{1}/\sqrt{2}\:$ \cite{Bravyi2005UniversalAncillas}, where the magic state can be produced with an error proportional to the physical error, independent of the code distance \cite{Litinski2019MagicThink}. To create a sufficiently high quality magic state, a distillation protocol \cite{Bravyi2012Magic-stateOverhead, Fowler2013SurfaceDistillation} can be used which essentially involves converting multiple low fidelity states into fewer higher fidelity states. Due to the high time cost associated with magic state distillation (the production and consumption), we can make a simplifying assumption that the time required to perform the T-gates effectively determines the final run time of an algorithm, as the relative cost of performing the Clifford gates fault-tolerantly is negligible. Some algorithms more naturally lend them selves to being expressed in terms of Toffoli gates, and there exist distinct specialized distillation factories to produce the magic states required to effectuate both of these non-Clifford operations. A Toffoli gate can be decomposed using 4 T gates \cite{Jones2013Low-overheadGate}, whereas the CCZ states normally produced for the Toffoli gate can be efficiently catalyzed into two T states \cite{Gidney2019EfficientTransformation}.   %To produce the following physical resource estimates we use lattice surgery based methods \cite{Litinski2019gameofsurfacecodes}, which among other improvements reduces the space-time cost of distillation protocols by up to 90\% in comparison to braiding-based implementations.
 
 \subsection{Error correction and logical error rate}\label{2.2}

A logical qubit in the surface code consists of data qubits, which store the quantum information, and ancillary qubits, which are used for stabilizer measurements that non-destructively extract error information from the data qubits. The distance, $d$, of a code represents the minimum size of physical error that can lead to a logical error and in the surface code the number of physical qubits per logical qubit scales as $2d^2$. The logical error rate per logical qubit, $p_L$, per code cycle as a function of the base physical error rate, $p$, can be approximated by \cite{Fowler2018LowSurgery}
\begin{equation}\label{eq2}
    p_L=0.1(100p)^{(d+1)/2}.
\end{equation} 

The efficiency of the error protection decreases as the base physical error rate approaches from below the threshold of the code (here assumed to be 1\%). For feasible final resource estimates the base physical error rate will need to be close to or below $10^{-3}$, where the necessary code distance is chosen based upon both the base physical error rate and the length of the desired computation. The physical error model here is the assumption that each physical operation has a probability $p$ of also introducing a random Pauli error. The achieved gate fidelity (the go-to metric for experimentalists) cannot be directly converted with confidence to the physical error rate, without further information. The cause of the gate infidelity, and where it exists on the spectrum between the two extremes of depolarizing error (decoherent noise) and unitary error (coherent noise) will determine the corresponding gate error rate. The best case situation is the one to one mapping between gate error (1-fidelity) and depolarizing error rate, and this has often been an assumption in the experimentally-focused literature. A measure of gate fidelity alone cannot determine the unitarity of the noise, i.e. the relative contribution of coherent and decoherent errors. Coherent errors, such as an over rotation of an intended gate, can positively interfere with each other and therefore cause worse case errors than those that are decoherent. The worst case scaling of the physical error rate, $p$, with gate fidelity $F$, and dimension of gate $D$, is $p=\sqrt{D(D+1)(1-F)}$\; \cite{Sanders2016BoundingFidelity}. To illustrate an example of this worst case scenario on noise quality, to guarantee an error rate of below 1\%, a gate fidelity of $99.9995\%$ would be required \cite{Sanders2016BoundingFidelity}. To determine where on the unitarity spectrum the actual hardware noise exists, protocols based on randomized bench marking can be used \cite{Wallman2015EstimatingNoise, Flammia2020EfficientChannels}. This information can then be used to estimate the base physical error rate with confidence \cite{Sanders2016BoundingFidelity,Kueng2016ComparingThreshold}. 

\subsection{Code cycle, reaction time and measurement depth}\label{2.3}
The code cycle is the base unit of operation in the surface code and it involves performing a full round of stabilizer measurements. As all operations in the computer are assumed to be subject to errors, including the stabilizer measurement process, the code cycle needs to be repeated d times before corrections can be applied. We will refer to the time it takes to perform these d rounds of code cycles as a ``\emph{beat}", where many surface code operations will have a time cost measured in \emph{beats}. A fault tolerant quantum computer using the surface code can be envisaged as partitioned into two sections, data-blocks which consume magic states to effectuate T gates for the desired algorithm, and distillation-blocks which produce high fidelity magic states. Each of these constructs in the surface code consist of a number of logical qubits, sometimes referred to as tiles, and each tile contains $2d^2$ physical qubits. The data block has a size scaling with the number of abstract qubits required for the algorithm and this then sets the minimum required number of physical qubits when paired with a single magic state factory and given the code distance. 

The T (or Toffoli) gates of an algorithm can be arranged into layers of gates (measurement layers), where all of the gates within a layer could potentially be executed simultaneously. Measurement layers are sometimes instead referred to as T layers when the relevant non-Clifford gate is the T gate, and the measurement (T) depth is the number of measurement layers in the algorithm. When a magic state is consumed by the data block, a Pauli product measurement is performed to determine whether an error has occurred so that a (Clifford) correction can be applied if required. The algorithm cannot proceed to the next measurement layer until all of the necessary corrections have been applied in the current layer, and this process requires a classical computation (decoding and feed-forward). The characteristic time cost that includes the quantum measurement, classical communication, and classical computation is referred to as the ``reaction time". It is conjectured that the fastest an error corrected quantum algorithm can run, i.e. the time optimal limit \cite{Fowler2012Time-optimalComputation}, is by spending only one reaction time per measurement layer, independent of code distance, and we will refer to this as reaction limited. In the case of superconducting devices which have relatively fast physical gates and measurements, the reaction time may be dominated by the classical communication and computation. A reaction time of 10 $\mu s$ has been used in recent resource estimation work \cite{Gidney2019HowQubits}, as compared to the 1 $\mu s$ code cycle time, which requires both physical two qubit operations and measurements. In this work we have defined the reaction time (RT) as a function of the code cycle time (CC) with $RT=(CC/4)+10\:\mu s$ unless otherwise stated. This assumption is motivated by the fact that generally the quantum measurement within the code cycle represents a fraction of the total time, for example with the shuttling based trapped ion architecture with a code cycle time of 235 $\mu s$, the quantum measurement is estimated to represent $\sim10\%$ of that total time. We include a code cycle independent additional cost of $10 \mu s$ to represent the (somewhat) hardware independent classical processing. With our resource estimation tool, which is available upon request, one could recreate the results of this paper with a different relationship between the code cycle time and reaction time. If particular surface code set up contained only a single distillation factory, then the rate of computation would likely not be limited by the reaction time but instead by the rate of magic state production. We refer to the regime of being limited by magic state production as ``\emph{tick}" limited. There are then three relevant regimes for surface code strategies which are separated by the limiting factor of computation rate. \emph{Beat} limited implies the limiting factor is the rate of magic state consumption by the data block, \emph{tick} limited, the rate of magic state production by the distillation blocks, and \emph{reaction} limited, where the conjectured time-optimal limit is reached and one reaction time is spent per measurement layer. 

In this work we utilize and compare two distinct strategies of incrementally trading qubits for run-time up to the reaction limit, and introduce them later in this section.

\subsection{Distillation and topological errors}

When choosing a surface code set up one must decide upon the acceptable final success probability, where in principle a success probability greater than $50\%$ would be sufficient to reach the desired precision by repeating the computation. The acceptable success probability then allows one an additional method of trading space for time, as the lower the acceptable success probability, the lower the various code distances would need to be. There are two contributions to the probability of failure, the topological error associated with the data block, and the total distillation error. The topological error is the chance for at least a single logical error within the data block across the entirety of the algorithm, which can be calculated with the product of the number of logical qubits, the number of code cycles required for the algorithm, and the logical error rate.  Where the logical error rate is defined in equation \ref{eq2} by the base physical error rate and the code distance on the data block. The total distillation error corresponds to the probability of at least a single faulty magic state which is calculated by the product of the total required number of required magic states and the error rate per state. The error rate per state is determined by the particular factory chosen and its associated code distances. We can then consider the final failure probability as the linear sum of the topological error and total distillation error. 

In this work we set the allowable total distillation error at $5\%$ implying that across the entire algorithm the probability that a magic state is generated with an error is $5\%$. We choose the appropriate distillation protocol to achieve this error rate, by selecting between T factory protocols of Litinski \cite{Litinski2019MagicThink}, and by adjusting the level 1 and level 2 code distances of the AutoCCZ factory. The choice of $5\%$ is in part motivated by the capacity of the AutoCCZ factory to reach a sufficient fidelity given the number of magic states required in the quantum advantage cases we address in this work.  We set the final topological error to be $1\%$ and choose the appropriate code distance to achieve this by considering the total number of code cycles that the algorithm must run for. This leads to a total final error (failure probability) of $\sim6\%$. In Litinski's work a final error of $2\%$ is chosen \cite{Litinski2019gameofsurfacecodes}, whereas in Gidney and Eker{\aa}'s work of breaking RSA encryption the final error is $33.4\%$.  The best choice of final error tolerance will depend on the type of problem being solved, in the case where the result being correct is heralded (e.g. factoring), one can accept large probabilities of failure, leading to a flexible trade-off between space and effective run time (including retries). Therefore in this case the choice should be framed as an optimization problem as opposed to an arbitrary threshold decision. Algorithms which require statistical accumulation from multiple runs may need the failure rate to be considerably lower than in the heralded type algorithms. Using our resource estimation tool, one could investigate the impact of different final failure probabilities.

\subsection{Routing at the error corrected level}\label{s2.4}
It is necessary to move logical information from one area of the device to another, and perhaps the most common requirement is the transport of magic states from the distillation factories to the data blocks. Analogous to physical qubits in superconducting devices which use logical operations to perform swaps, one can imagine performing swap operations between logical qubits in the surface code. However, alternative techniques are more often considered for enabling long range interactions with topological error correcting codes. Using lattice surgery based methods, long range interactions (CNOTs) can be enabled in d code cycles (1 \emph{beat}), essentially independent of the distance between the two points, so long as there is a chain of free ancillary qubits between them. We refer to this method as entanglement swapping. An ancillary logical qubit can only contribute to one routing chain at a time (per beat).  When defining the layout of an error correction set up, which involves choosing the number of distillation factories and orienting them with respect to the data blocks, one must also ensure there is sufficient ancillary routing space to enable the required degree of data transfer between the relevant areas. The degree of data transfer, or alternatively stated the degree of parallelization in the execution of the algorithm, is often characterized by the number of magic states consumed per beat across the entire data block. With a data block arranged into rows with an access hallway between each row (consisting of ancillary logical qubits), there are two unique ways of touching each data qubit, if this is deemed insufficient, entangled copies of the data rows can be created to ensure there are enough unique paths between the factories and the data qubits. In section \ref{s2.7} we present our choice of the routing overhead factor for entanglement swapping as a function of the degree of parallelization using AutoCCZ factories.

\subsection{Considering physical mid-range connectivity}\label{s2.6}

In this section we briefly consider the potential impacts on the required resources when the underlying architecture has access to flexible mid-range connectivity between physical qubits. For superconducting devices two qubit operations can only be performed between nearest neighbour physical qubits, and so long range interactions must be enabled by sequences of costly logical swap operations. Alternative hardware may have access to low overhead long range interactions between physical qubits, for example, photons are readily transported long distances which is relevant to both photon-only hardware, and for connecting small modules of trapped ions with photonic interconnects. In the blueprint for a shuttling based trapped ion architecture \cite{Lekitsch2017}, a single system is envisaged, comprised of iterable micro-fabricated chips, connected via electric fields, which allow for physical shuttling between modules. High state fidelity adiabatic shuttling has been demonstrated with a speed of $\sim 20$ $ms^{-1}$ \cite{Kaufmann2018}; diabatic techniques \cite{Torrontegui2011FastHeating, Walther2012ControllingIons, Bowler2012CoherentArray} can enable much greater shuttling speeds, and $\sim 80$ $ms^{-1}$ has been demonstrated \cite{Walther2012ControllingIons}. Lau and James calculate that the maximum speed a $^{40}Ca^+$ ion can be transported across a 100 $\mu m$ trap without excessive error is 10,000 $ms^{-1}$ \cite{Lau2011DecoherenceTransport}. For NISQ size devices all to all connectivity should be achievable with high fidelity (relative to two qubit operations) using the shuttling-only approach \cite{Webber2020EfficientComputer}, but physically shuttling completely across a device with over a million physical qubits is unlikely to be feasible due to the associated time cost. Utilization of mid-range connectivity may still enable a reduction in the routing overhead associated with entanglement swapping.

While very long range shuttling operations may be protected from error by periodic cooling operations and mid-circuit syndrome extraction and correction, the total time cost must be considered. With entanglement swapping, long range interactions can be enabled between logical qubits in the surface code in a single beat, provided there are sufficient available ancilla qubits between the locations. To contrast this capability we estimate the range at which physical shuttling may remain competitive with entanglement swapping. Assuming a code distance of 30, and logical qubits distributed across a 2D square grid, we estimate that a logical qubit qubit could interact via physical shuttling with another logical qubit in the range of 3-30 grid spaces away within a single beat (d code cycles), depending on physical ion density and shuttling speed. While this is indeed unlikely to be sufficient for mediating all long range interactions between logical qubits, the capability of low cost mid-range physical connectivity could make the transversal CNOT preferable to the more usually considered lattice surgery based methods. Guti\'errez et al have investigated the experimental regimes at which the transversal CNOT may outperform the lattice surgery based methods for trapped ions \cite{Gutierrez2019TransversalityProcessors}, and for particular values of error contributions, the transversal CNOT may be performed a factor $10\times$ faster. If the transversal CNOT is expected to require less time than the lattice surgery based CNOT, then this would directly impact the rate of magic state production of distillation factories and therefore could reduce the total qubit overhead in the highly parallelized regime. Furthermore, mid range physical connectivity could considerably reduce the qubit footprint of distillation factories by eliminating the need for interior ancillary routing space for entanglement swapping. Alternative error correction strategies to the 2D surface code \cite{roffe2020decoding, ScrubyNumericalCode} may be achievable on hardware with flexible mid-range connectivity. We leave a more detailed analysis of the potential benefit of mid-range connectivity as future work. 

In the next subsection we introduce the Game of Surface Codes method of trading space for time \cite{Litinski2019gameofsurfacecodes}, and following that, the AutoCCZ method \cite{Gidney2019HowQubits} where we also include some more detailed assumptions on the necessary routing overhead for entanglement swapping. 

\subsection{Game of Surface Codes}
In the work of Litinski \cite{Litinski2019gameofsurfacecodes}, various data blocks are presented which vary in their rate of T-gate effectuation (magic state consumption) and the number of required physical qubits. There are numerous distillation protocols each of which varies with regard to the output fidelity, required number of physical qubits, and the rate of production. Distillation blocks can be parallelized and this enables further space-time trade-offs. 

We make use of distillation strategies presented by Litinski \cite{Litinski2019MagicThink} where the distance associated with the distillation blocks is fine tuned and separate from the distance associated with the data blocks. The data blocks have a required code distance set by the total number of logical qubits and the total number of code cycles required to run the entire algorithm. With the total number of T gates in the algorithm, $T_{count}$, the distillation blocks only need a code distance sufficient to produce magic states with an error lower than $1/T_{count}$. The distillation blocks use a certain number of qubits and only need to be protected for a certain number of code cycles (corresponding to one full round of distillation), and these are generally both small relative to the requirements of the data blocks. This method of individual calibration of distance for the data and distillation blocks is in contrast to a prior method \cite{Litinski2019gameofsurfacecodes} where both block types are attributed the same code distance.

In the GoSC scheme, Clifford gates are addressed explicitly via Clifford tracking, i.e. all Clifford gates are commuted to the end of the circuit and absorbed into measurements. This turns T gates into Pauli product rotations, and measurements into Pauli product measurements. In general these Pauli product rotations can be big multi-qubit operations. To account for the generalized worst-case algorithm, which may have these multi-qubit operations, the maximum rate at which the data block can consume a magic state is defined as one state per \emph{beat} (d code cycles). If the input circuit is known then it will sometimes be possible to arrange the data blocks in such a way so that more than one state can be consumed per \emph{beat}. In this investigation we do not consider the details of the input circuit and instead rely only meta details such as, abstract qubit count, total T count, and the measurement depth. Our utilization of the GoSC method should then be considered an upper-bound configuration, i.e.,  guaranteed to be able to support any algorithm configuration given the meta details. The number of distillation factories can be chosen to match the production rate to the consumption rate of the data block, at which point we may describe the strategy as \emph{beat} limited. 

In the work of Litinski \cite{Litinski2019gameofsurfacecodes} further time optimizations are presented, which can be utilized to reach reaction limit where one reaction time is spent per measurement (T) layer. The reaction limited strategy is set by the total number of measurement layers, i.e. the measurement depth ($T_{depth}$), as opposed to the total T gate count ($T_{count}$) of the \emph{beat} limited strategy. The average number of parallel-executable T gates per layer ($T_{layer}=T_{count} / T_{depth}$) varies across particular algorithms and there exist methods to optimize circuits to minimize either the $T_{count}$, $T_{depth}$, or the circuit width \cite{Amy2014Polynomial-timePartitioning,AbdessaiedTechnologyCircuits}. Litinski's method of speeding computation up beyond the \emph{beat} limited case utilizes ``\emph{units}" which combine the previous constructions of data and distillation blocks. The number of units can be increased until the time optimal limit (henceforth reaction-limited) is reached, where each unit can work in parallel to address a set of measurement layers. This does not contradict the previously provided definition of measurement layers, as although the units can parallelize aspects of the work, the layers must still be stitched back together requiring one reaction time per measurement layer for corrections. The reaction limit then defines the maximum number of units that can be utilized. For an algorithm requiring $n$ abstract (logical) qubits, with an average of, $T_{layer}$, T gates per measurement layer, a single unit will consist of $4n +4\sqrt{n}+1$ tiles (logical qubits) for the data block, and $2\;T_{layer}$ storage tiles. Each unit requires an amount of time, $t_u$, to process a single measurement layer, where it is measured in \emph{beats} and scales as $T_{layer} + \sqrt{n} +3$. Each unit will need a number of distillation factories to match the required production rate, which is $T_{layer}$ magic states per unit completion time, $t_u$. With a linear arrangement of units as is considered here, and a reaction time, $RT$, the reaction (time-optimal) limit is reached with a number of units $n_u$ equal to $t_u/RT+1$. The GoSC scheme of going beyond the \emph{beat} limited rate initially incurs a space-time overhead but then allows one to trade linearly (by increasing the number of units) until the reaction limit is reached.

\subsection{AutoCCZ factories} \label{s2.7}

In the work of Gidney and Eker{\aa} \cite{Gidney2019HowQubits} detailed surface code layouts along with the logical algorithmic developments for breaking RSA encryption are provided. The surface code strategy utilizes AutoCCZ factories \cite{Gidney2019FlexibleStates} which create a magic state that can be consumed to effectuate the non-Clifford Toffoli gate. The ``Auto" refers to the fact that these factories create auto corrected CCZ states, meaning that the potential correction operation associated with magic state consumption is decoupled and can be performed far away from the data block.

Using this scheme with the AutoCCZ factories, the \emph{beat} limited rate can be surpassed without introducing units as is performed in GoSC. Provided the routing overhead is accounted for, one can continue to add AutoCCZ factories until the production rate is equal to the number of Toffoli gates per measurement layer per reaction time, at which point the reaction limit is reached. 

%Only a single data block is required in the AutoCCZ scheme, with size scaling linearly with the number of abstract qubits. This is in stark contrast to the GoSC method where every unit added requires its own comparably sized data block; in figure \ref{fig:1} we compare the final qubit overhead between these two methods.

The AutoCCZ factory is characterized by two code distances, corresponding to the two levels of the protocol. The factory is a tiered distillation scheme where the output magic states of the first level are the input states to the second level. We calculate the final output fidelity following the ancillary files of Gidney and Eker{\aa} \cite{Gidney2019HowQubits} as a function of the two code distances and base physical error rate. To choose the optimal two code distances we assess a wide range of possible values and select the setup that reaches the desired final distillation error rate while minimizing the space-time volume of the factory. While technically possible to maintain a reasonably low distillation error rate with base physical error rates near the threshold of the code, $10^{-2}$, the associated code distances required would result in an infeasible physical qubit overhead. Assuming any code distance is acceptable, the final distillation error per state, $p_D$, can be described as a function of the base logical error, $p$, by $p_D=34300p^6$, until the threshold of the code is reached. This relationship was found by numerical fitting. A limit on the allowable code distance breaks away from this trend before the threshold, and the lower the limit the sooner it breaks upwards. In figure \ref{fig:2}B we investigate the impact of the base physical error rate for the final qubit overhead to reach a desirable run time with the AutoCCZ method. If the final output fidelity of the AutoCCZ factory is insufficient given the desired length of an algorithm and base physical error rate, then alternative T gate factories may still be viable \cite{Litinski2019MagicThink}.

To reach a particular desired run time, assuming it is below the reaction limit, the number of AutoCCZ factories are chosen as necessary. Once this footprint arrangement is settled (the combination of the data block and the number of factories), the last stage of the calculation is to determine the necessary routing overhead to account for the degree of parallelization. As described in section \ref{s2.4}, long distance interactions between the distillation block and data block can be enabled in one beat provided there is an available chain of ancillary logical qubits (routing space) between them. As each ancillary logical qubit can only contribute to one routing chain per beat, there may need to be additional unique paths to account for the degree of parallelization. All of the necessary routing overhead is strictly accounted for in GoSC, with the construction and arrangement of the data blocks, which are then duplicated and distributed across units. In our utilization of AutoCCZ factories we define the degree of parallelization as the number of magic states consumed across the data block per beat. We then ensure there is a routing hallway per data block row for every state consumed per beat, to exceed two hallways per data row, entangled copies of the data block can be made (which is comparable to the entangled copies across Units in GoSC). Finally we multiply the entire area (number of logical qubits) by a factor of 1.2 to ensure the distillation factories are surrounded by hallways and for some work space around the data block for arranging routing-chains. Our utilization of AutoCCZ factories should not be considered a true upper bound for any generalized circuit, unlike GoSC Units. In the following section we contrast the two methods further and state our relative contribution.

\subsection{Problem specification} \label{s2.8}
We start our investigation with the logical resource requirement set out by Lee et al \cite{Lee2020EvenHypercontraction} to simulate FeMoco to chemical accuracy, and assess the feasibility of reaching desirable run times as a function of the code cycle time and number of achievable physical qubits. In the work of Lee et al, a detailed surface code strategy is presented with algorithm specific optimizations, such as minimizing the number of data qubits that are stored for working. In contrast, the surface code strategies we utilize, do not necessarily require detailed knowledge of the input circuit, and are instead only a function of the logical qubit count, the T (or Toffoli) count and the measurement depth. This approach may not yield optimal results for specific algorithms but it enables one to effectively estimate physical resource requirements for a particular algorithm as a function of strategy (the time-space optimization spectrum) and the code cycle time. We contrast two strategies, the first which can be considered to provide upperbound resource estimates for any general circuit input, and closely follows the work of Litinski's Game of Surface Codes (GoSC) \cite{Litinski2019gameofsurfacecodes}. We go beyond the pedagogical examples provided by Litinski \cite{Litinski2019gameofsurfacecodes} by creating an automatic tool that calculates the physical resources across the space-time optimization spectrum and apply it to the algorithmic requirements of quantum advantage use cases. Furthermore we use the tool to calculate the number of physical qubits required to reach a specified run time as a function of the code cycle time of the hardware by utilizing the necessary degree of parallelization. We investigate the impact of varying the measurement depth with a fixed total T count, on the efficiency of the GoSC unit approach and identify the optimal value in particular regimes.   In addition to the Game of Surface Codes method of parallelization, we incorporate a method which uses AutoCCZ factories \cite{Gidney2019FlexibleStates}. For this method we relied upon the ancillary files of Gidney and Eker{\aa} \cite{Gidney2019HowQubits} and adapted them to be flexible enough for our broader (circuit agnostic) considerations. Our intention with the routing overhead with the AutoCCZ factories is to cover a wide range of possible circuit characteristics, while it may be an over estimation for some specific circuits, it may also represent an under estimation for some worst-case situations. We accomplish our aim of quick and general resource estimation as a function of algorithm meta information, and hardware characteristics, by contrasting the upper-bound scenario of GoSC units, with a more heuristic utilization of AutoCCZ factories. With detailed knowledge of the input circuit, further optimizations of the footprint configuration are possible, but generally these must be performed on a case by case basis and are non trivial to automate. The tool used to generate the results presented in this paper is available upon request. We use the latest logical resource requirements for breaking elliptic curve encryption \cite{HanerImprovedLogarithms,RoettelerQuantumLogarithms} and estimate the number of physical qubits required to break the encryption of Bitcoin keys in the small amount of time it would actually pose a threat to do so, all as a function of both the code cycle time and base physical error rate.

\begin{figure}[t!]
\centering
\includegraphics[width=17.6cm]{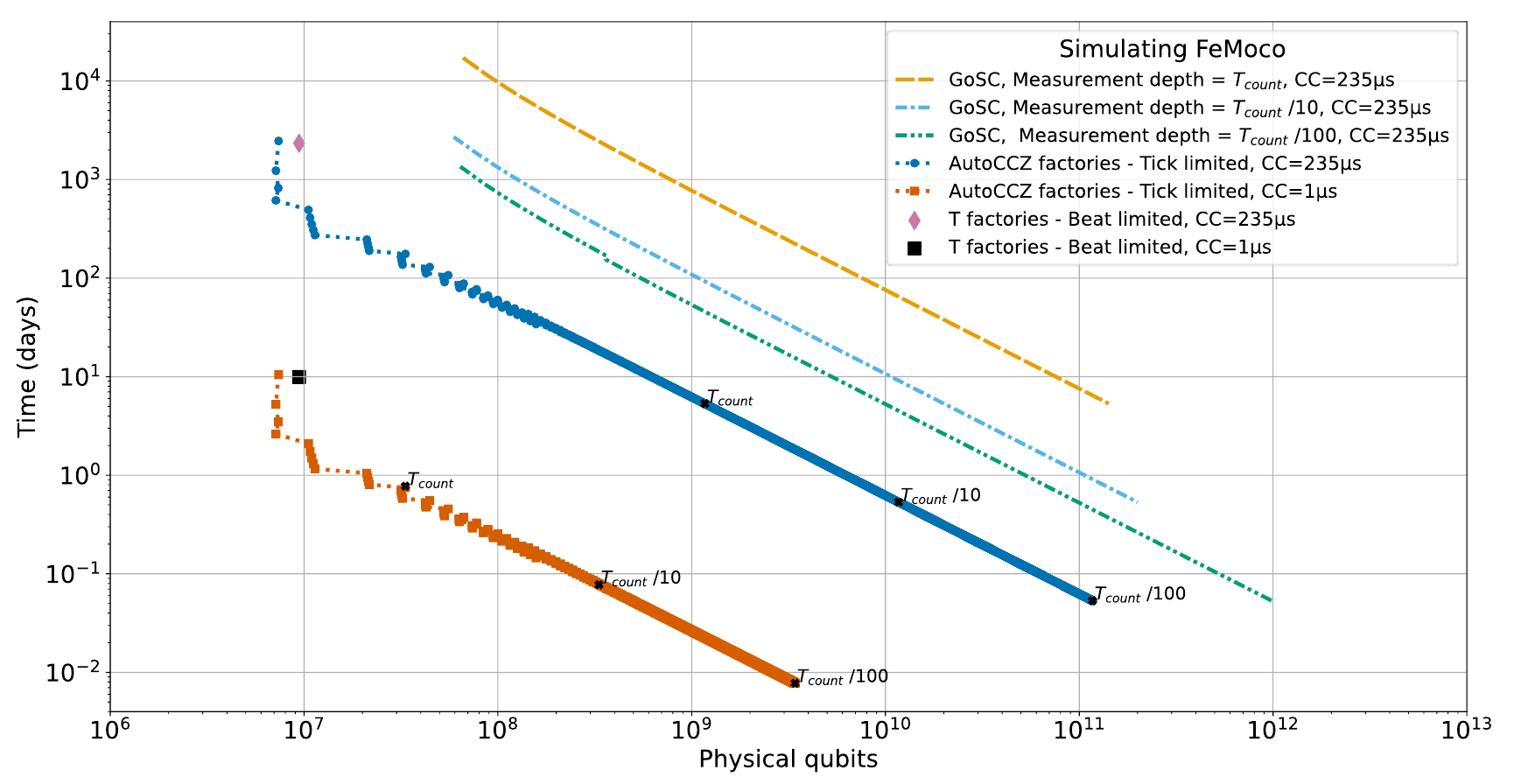}
\caption[Tof gate cost]{\label{fig:1} Physical resource estimates for a ground state energy calculation of the FeMoco molecule to chemical accuracy as per the logical resource requirements of Lee et al \cite{Lee2020EvenHypercontraction} which makes use of Tensor Hypercontraction (THC) and the Qubitization method. The associated logical resources required are 2196 logical qubits and 6.7 billion Toffoli gates. Here the run time (in days) and associated number of physical qubits required for the energy calculation are plotted for different surface code strategies and code cycle times. A code cycle time of 1 $\mu s$ and reaction time of 10 $\mu s$ is considered, which may correspond to future superconducting devices, in addition to a code cycle time of 235 $\mu s$ with reaction time of 70 $\mu s$ which may correspond to a future shuttling based trapped ion architecture \cite{Lekitsch2017}. The base physical error rate is set to $10^{-3}$, the final distillation error probability is at most $5\%$, and the final topological error probability is at most $1\%$. The method from A Game of Surface Codes (GoSC) \cite{Litinski2019gameofsurfacecodes} is utilized where layers of T gates are parallelized using ``Units" and is shown for the code cycle time of 235 $\mu s$. The three distinct trends all with dashed lines correspond to different measurement depths given as a fraction of total non-Clifford gate count, $T_{count}$. The start of the trends represent the minimal number of units with the linear arrangement, which is 3, and then units are sequentially added up until the reaction limit is reached. With the GoSC units approach, the run time for a code cycle of 1 $\mu s$  for a given physical qubit number can be inferred from the 235 $\mu s$ case by scaling by the same relative difference in the code cycle time. A \emph{beat} limited approach (i.e. limited by the magic state consumption rate of the single data block) through the GoSC lens are plotted as a square and diamond. With the GoSC approach the improved distance T gate factories of Litinski \cite{Litinski2019gameofsurfacecodes} were used, where their distance is calibrated separately to the data blocks, and it is assumed that 4 T gates can be used to construct a Toffoli gate. A distinct method of parallelization is utilized here, which uses AutoCCZ factories \cite{Gidney2019FlexibleStates, Gidney2019HowQubits}, enabling all of the Toffoli gates within a given measurement layer to be potentially performed in parallel. This approach is referred to as \emph{tick} limited implying that it is the rate of magic state production that limits the computation rate, up until the reaction limit is reached.  The AutoCCZ approach is plotted for the two code cycle times of 1 $\mu s$ and 235 $\mu s$ as joined markers where the trend starts with 1 AutoCCZ factory and sequentially adds factories, each represented by a data point, up until the reaction limit is reached. The reaction limit for a given measurement depth is plotted and is labeled with the measurement depth expressed as a fraction of the total non-Clifford gate count, $T_{count}$, $T_{count}$/10, and $T_{count}$/100. }
\end{figure}

\section{Results}\label{s3}
To calculate the results presented in this section we use various surface code strategies including the Game of Surface Codes scheme which uses units to parallelize layers of T gates \cite{Litinski2019gameofsurfacecodes}, and AutoCCZ factories \cite{Gidney2019FlexibleStates,Gidney2019HowQubits}, which are both highlighted in the previous section.

\subsection{Simulating FeMoco as a function of the code cycle time}\label{3.1}

There has been extensive research into both algorithmic development and resource estimation in the field of fault tolerant quantum chemistry, and one of the focus points has been the FeMo-co catalyst  \cite{Reiher2017, Li2018TheSimulations, vonBurg2020QuantumCatalysis,Lee2020EvenHypercontraction}. An improved understanding of the FeMo-co catalyst could provide considerable efficiency improvements to nitrogen fixation which currently represents around $2\%$ of the worlds energy usage. We start with some of the latest algorithmic developments by Lee et al \cite{Lee2020EvenHypercontraction} and investigate the feasibility of achieving a reasonable run time for different code cycle times and different surface code strategies. The associated logical resources required are 2196 logical qubits and 6.7 billion Toffoli gates.

In figure \ref{fig:1} we compare two distinct methods of trading space for run time up to the reaction limit (the conjectured time optimal limit \cite{Fowler2012Time-optimalComputation}). The two scatter trends utilize AutoCCZ factories with different code cycle times of 1 $\mu s$, corresponding to a future superconducting device, and 235 $\mu s$, corresponding to a future shuttling based trapped ion device \cite{Lekitsch2017}. Each trend starts with one AutoCCZ factory, from there the number of factories is incremented and at each step the code distance is calibrated and the resulting run time and physical qubits are plotted. Initially the total qubit footprint is dominated by the data blocks whereas the run time is bottle necked by the magic state production, i.e. going from one factory to two halves the expected run time. Therefore while this is the case, adding factories results in an improvement to the space-time volume and this can sometimes allow for a reduction in code distance, which may result in an actual reduction in total qubit count. Each time a factory is added, the magic state consumption rate per beat is defined to determine whether the routing overhead needs to be increased as described in section \ref{s2.7}.  What is considered to be a desirable run time will largely depend on the importance of the problem being solved and by the speed and quality of the classical alternatives. With one AutoCCZ factory the superconducting device completes in around 10 days with 7.5 million qubits, whereas the trapped ion device requires ~2450 days and the same number of qubits. Where 10 days may be considered a quantum advantage for this use case where classical computers stand no chance of providing a meaningful answer, perhaps 2450 days would not. By parallelizing the magic state production the trapped ion device can reach the run time of 10 days requiring 600 million qubits. The factor difference between the physical qubit count here is less than the factor difference between the code cycle time, because initially adding factories is a favourable space time trade until the total qubit footprint becomes dominated by the factories at which point the trade becomes linear. The good news for hardware with slower code-cycle times is that it will often be possible to still reach desirable run times provided enough physical qubits are available. However, the associated qubit overhead may appear daunting, and implies that hardware with slower code cycle times will have to be more scalable to compete, assuming equal error rates and physical connectivity. We plot for a range of possible measurement depths, labeled as a fraction of the total Toffoli count, as this was not provided along with the other logical requirements \cite{Lee2020EvenHypercontraction}. In the AutoCCZ scheme the measurement depth does not directly impact the efficiency of the approach, instead it only determines in combination with the reaction time, what the time optimal (reaction) limit is. The labels then indicate the reaction limit, the point at which the trend would end, given that measurement depth. 

The assumption of a base physical error rate of $10^{-3}$ is often made in the literature \cite{Gidney2019HowQubits} and may be representative of two qubit gate fidelities that have already been achieved experimentally, depending on the potential caveats highlighted in section \ref{2.2}. Trapped ion architectures have achieved some of the highest gate fidelities to date \cite{Srinivas2021High-fidelityQubits, Schafer2018, Ballance2016, Gaebler2016High-FidelityQubits, Webb2018}, and generally exceed that of superconducting devices \cite{Zhang2020High-fidelityQubits, Xu2020High-FidelityQubits}, for that reason, here we provide the resource estimate for the more optimistic value of $10^{-4}$ for the base physical error rate, paired with the code cycle time that may correspond to future trapped ion devices. Instead of the 600 million physical qubits that was required for a run time of 10 days, now with a base physical error rate of $10^{-4}$  only 60 million physical qubits would be required. Although here the factor improvement of the base physical error rate is approximately equal to the factor improvement to the physical qubit requirement, this will not always be the case. The factor difference will depend on the proximity to the threshold of the code, and we investigate the impact of the base physical error in more detail in figure \ref{fig:2}B.

In figure \ref{fig:1} we also include the Game of Surface Codes (GoSC) approach to trading space for time \cite{Litinski2019gameofsurfacecodes}, where measurement layers are parallelized with constructs called ``Units". Each unit contains its own copy of the data block and enough factories to produce the number of magic states within the measurement layer within the time it takes to prepare the unit, which scales with both the number of magic states per layer and the number of abstract qubits. Units can be incrementally added, each one added reduces the final run time up until the reaction limit is reached. The dashed lines in figure \ref{fig:1} use units along with improved T gate factories \cite{Litinski2019MagicThink} where it is assumed 4 T gates are required to decompose a Toffoli gate \cite{Jones2013Low-overheadGate,Litinski2019gameofsurfacecodes}. The GoSC approach is plotted only for the code cycle time of 235 $\mu s$ but three trends are included for the different measurement depths as a fraction of the total Toffoli count. The efficiency of the GoSC approach (in addition to the reaction limit) is dependent on the measurement depth as can be seen. In section \ref{3.3} we investigate the impact of the measurement depth on the final qubit requirement to reach a fixed run time for the GoSC approach.

It can be seen in the figure that the AutoCCZ approach provides more favourable final resource estimates than GoSC units in this scenario. The largest contributing factor to this difference appears to be the initial set up cost for the unit approach, which is nearly an order of magnitude increase in qubits for no appreciable speed up (with measurement depth = $T_{count}/10$). This is in stark contrast to the AutoCCZ approach which initially takes very favourable space-time trades by increasing the number of factories. As the rate of parallelization increases in the AutoCCZ approach, eventually entangled copies of the data block are made to maintain sufficient access hallways between the data block and distillation blocks. Both the AutoCCZ and GoSC approach converge towards an equal linear trade between space and time. It should be restated that the GoSC approach can be considered a true upper-bound estimate, functional for any general circuit, whereas our utilization of the AutoCCZ factories is more heuristic and may represent an underestimation for some specific circuit inputs, see section \ref{s2.8} for more discussion on this.

In figure \ref{fig:1} a \emph{beat} limited method is included for both code cycle times as diamond and square points i.e. no units and the computation rate is limited by the data blocks magic state consumption rate. These points use T gate factories and the fast data blocks from GoSC, which have size scaling as $2n +\sqrt{8n}+1$ for n logical qubits, and can effectuate T gates at a maximum rate of one per \emph{beat} (d code cycles). The number of T factories is chosen to match the rate of magic state consumption of the fast data blocks. The \emph{beat} limited situation provides comparable run times to the single AutoCCZ factory and requires a similar number of physical qubits. 

To conclude this section, it appears that the AutoCCZ factories are the favourable approach to trade space for time up to the reaction limit, but a more detailed investigation into the underlying assumptions of both methods is warranted. These resource estimates are solely a function of algorithm meta information such as total T count, measurement depth, and the number of logical qubits. The surface code configuration can be optimized when paired with detailed knowledge of the input algorithm, but this process is non trivial to automate, as we have done in this work as a function of code cycle time. Future hardware that expects to have considerably slower code cycle times than the superconducting devices may still be able to reach desirable run times provided enough physical qubits are available, which further emphasises the importance of scalability. The associated qubit overhead factor will range between less than 1 (here $\sim 0.3$), to 1, times the difference in the code cycle time depending on the relative degree of parallelization in the comparison. Algorithms should be optimized by minimizing the measurement depth if the reaction limit is restrictive. The physical qubit requirement may be reduced if the underlying hardware has access to low overhead mid-range physical connectivity, as discussed in section \ref{s2.6}

\subsection{Breaking Bitcoin's EC encryption}\label{3.2}

\begin{figure}[t!]
\centering
\includegraphics[width=17.6cm]{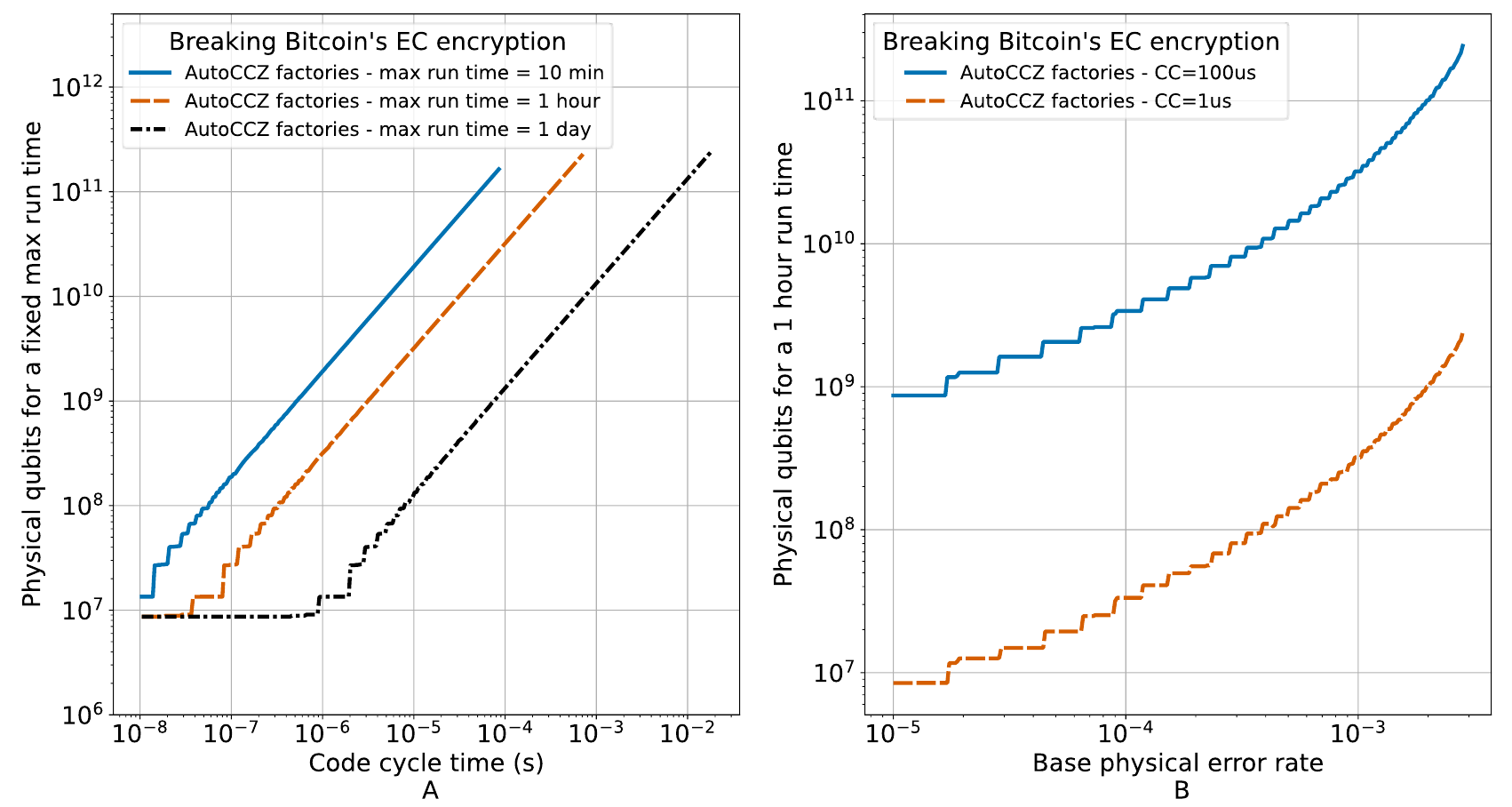}
\caption[Tof gate cost]{\label{fig:2} The number of physical qubits required to break Bitcoin's 256 elliptic curve encryption with a fixed maximum run time as a function of the code cycle time and base physical error rate. Using the latest algorithmic development for quantum circuits for elliptic curve encryption \cite{HanerImprovedLogarithms}; the depth optimized approach is chosen requiring $5.76\cdot 10^{9}$ T gates, 2871 logical qubits, and a T (measurement) depth of $1.88\cdot 10^{7}$. These trends utilize AutoCCZ factories to trade space for time to reach the desired run time and assume that a CCZ state can be efficiently traded for 2 T gates \cite{Gidney2019EfficientTransformation}. Assuming the relationship between reaction time (RT) and code cycle time (CC) of $RT=CC/4+10\;\mu s$ which is motivated in section \ref{s2}. (A) Physical qubit requirement as a function of code cycle time for maximum run times of 10 minutes, 1 hour, and 1 day.  The trends start from the chosen code cycle time of $10^{-8}$ and end at the right due to reaching the reaction limit before reaching the desired run time. We use a base physical error rate of $10^{-3}$ and a final output error of $\sim6\%$. (B) Physical qubit requirement as a function of the base physical error rate for a maximum run time of 1 hour and code cycle time of both 1 $\mu s$ and 100 $\mu s$. The trends start from the chosen base physical error rate of $10^{-5}$ and end at the right due to the AutoCCZ factory being no longer able to reach the desired distillation fidelity given the base physical error and number of required states. }
\end{figure}

Breaking encryption has received a lot of attention in the quantum computing community since Shor's breakthrough algorithm \cite{Shor2002AlgorithmsFactoring} which provides a near exponential speedup for prime factoring which has direct implications for breaking RSA encryption. Gidney and Eker{\aa} provide algorithmic improvements in addition to the surface code strategies for breaking RSA encryption and they estimate that 20 million qubits running for 8 hours could break it with a code cycle time of 1 $\mu s$ \cite{Gidney2019HowQubits}. In a blueprint for a shuttling based trapped ion device, which estimated the code cycle time to be 235 $\mu s$ \cite{Lekitsch2017}, it was originally estimated that breaking RSA encryption would require 110 days and 2 billion qubits with a base error of $10^{-3}$, implying a device occupying an area of $103.5 \times 103.5$ m$^2$. With the latest algorithmic and surface code strategy improvements we can reduce this estimate to requiring instead a run time of 10 days (a factor $10\times$ faster) and 650 million qubits, which would imply a device size of area $60 \times 60$ m$^2$. With a base physical error of $10^{-4}$ the device size would reduce to $18 \times 18$ m$^2$ and further reductions may be possible if one were to make use of the flexible mid-range connectivity that is available. Using the same relative improvement factor from mid-range connectivity as in the design blueprint \cite{Lekitsch2017}, starting from the $10^{-4}$ case, we might expect a device of size $2.5 \times 2.5$ m$^2$ to be sufficient. This is indeed a rough estimate and a more rigorous understanding of how to best make use of mid-range connectivity would be required to confidently provide the performance improvements relative to a nearest neighbour approach. 

Bitcoin uses the Elliptic Curve Digital Signature Algorithm (ECDSA) which relies on the hardness of the Elliptic Curve Discrete Log Problem (ECDLP) and a modified version of Shor's algorithm \cite{Shor,HanerImprovedLogarithms, RoettelerQuantumLogarithms} can provide an exponential speedup using a quantum computer for solving this problem. The encryption of keys in the Bitcoin network are only vulnerable for a short window of time, around 10 minutes to an hour depending on the fee paid, as described in more detail in section \ref{1.2}, and this makes it a well suited problem for our investigation. In the work of Aggarwal et al \cite{AggarwalQuantumThem}, the potential threat of a quantum attack on the Bitcoin network is investigated, and two main vulnerabilities are contrasted, i.e., proof of work (mining) and the encryption of private keys. Following the conclusions of Aggarwal et al, we highlight the differences between these two vulnerabilities and the associated quantum speed up in section \ref{1.2}. Here we focus on the most likely threat, the encryption of private keys, and go beyond the physical resource estimates of Aggarwal by including the latest improved logical algorithmic requirements of H\"{a}ner et al \cite{HanerImprovedLogarithms}, and by quantifying how the number of physical qubits changes based on the desired run time and the code cycle time of the hardware. In the following we will present our resource estimates for this problem, and also contrast with the work of Aggarwal et al \cite{AggarwalQuantumThem}.

In figure \ref{fig:2}A we plot the number of physical qubits required to break the elliptic curve encryption of Bitcoin within a run time of 1 day, 1 hour and 10 minutes, as a function of the code cycle time. We use the logical resources of the depth optimized approach provided by Haner et al  \cite{HanerImprovedLogarithms} for 256 bit encryption, which corresponds to $5.76\cdot10^9$ T gates, 2871 logical qubits, and a measurement depth ($T_{depth}$) of $1.88\cdot10^7$. The measurement depth of this algorithm is low relative to the $T_{count}$, at $\sim T_{count}/300$, implying that there is a lot of room for parallelization before the reaction limit is reached. In figure \ref{fig:2}A it can be seen that it would require 317 million physical qubits to break the encryption within one hour with a code cycle time of 1 $\mu s$. To break it within 10 minutes with the same code cycle time it would require 1.9 billion physical qubits whereas to break it in 1 day would require only 13 million physical qubits. The horizontal period most evident in the dash-dotted black trend for a run time of 1 day is because the desirable run time can reached for those code cycle times with only 1 AutoCCZ factory. Once the code cycle time increases sufficiently it is then necessary to begin adding AutoCCZ factories to maintain the desired run time. Hardware with considerably slower code cycle times than 1 $\mu s$ will need to be able to reach larger device sizes to break the encryption within the allotted time. Even for code cycle times of 1 $\mu s$, this large physical qubit requirement implies that the Bitcoin network will be secure from quantum computing attacks for many years. High value transactions are likely to pay high fees ensuring they are processed with higher priority, and therefore would require considerably more physical qubits to break the encryption in time. The Bitcoin network could nullify this threat by performing a soft fork onto an encryption method that is quantum secure, where Lamport signatures \cite{Abdullah2018AdoptionIoT} are the front-running candidate, but such a scheme would require much more memory per key. The bandwidth of Bitcoin is one of the main limiting factors in scaling the network and so changing the encryption method in this way could have serious drawbacks.

The logical resources provided by H\"{a}ner et al \cite{HanerImprovedLogarithms} for breaking elliptic curve encryption improve on the prior state of the art of Roetteler et al \cite{RoettelerQuantumLogarithms} by over an order of magnitude. In the quantum threat to Bitcoin work of Aggarawal et al \cite{AggarwalQuantumThem} the older and less favourable logical resource requirements were considered \cite{RoettelerQuantumLogarithms}. Aggarawal et al estimate it would require 6.5 days and 1.7 million physical qubits to break the encryption with a base physical error rate of $5\times10^{-4}$. The code cycle time is not explicitly defined, and instead a physical gate rate of 66.6 MHz is assumed, which we estimate would correspond to a code cycle time of approximately 0.1 $\mu s$. Next we calculate the physical resources using the assumptions and logical requirements of Aggarawal et al, and we find that a device with 3 AutoCCZ factories, would complete in 7 days and require 5 million physical qubits. There is rough agreement between the final physical resources between our methods; the remaining discrepancy originates from the differing estimates of the number of physical qubits that are required per logical (abstract) qubit.  The conversion factor between logical to physical qubits of Aggarawal et al is stated to be 735.5 for this problem, which should include the overhead associated with the degree of encoding (code distance), distillation factories, and routing space. We find that a code distance, d, of 25 is required to maintain a final failure rate below $6\%$, implying at least $2\times d^2$, or 1250, physical qubits per logical qubit. When we include the distillation and routing overhead our final physical to logical qubit conversion factor is 2140. While a higher final failure rate may be tolerable in a heralded problem such as breaking encryption, we estimate that a code distance of at least 22 would be required, as a code distance of 21 or lower would lead to final failure rates of $\sim100\%$.

H\"{a}ner et al \cite{HanerImprovedLogarithms} provide asymptotic formulas for the resources required to break the elliptic curve encryption as a function of the input encryption bit size, n (Bitcoin uses 256 bit encryption). The total logical space-time volume (T count $\times$ logical qubits) scales to leading order with encryption bit size as $\mathcal{O}({n^4}/\log{n})$, which can be compared to the exponential time complexity of classical methods, $\mathcal{O}(2^{\sqrt{n}})$. We calculated the physical resources to break the elliptic curve encryption as a function of bit length and one may expect the results to follow the same form of dependence as the total logical space-time volume. This is because we would expect the total physical qubits required to scale linearly with the total logical space-time volume, and this is generally the case except for variations due to differing efficiencies of the error correction set up at different regimes (ratio of the logical qubits to gate requirement, and the degree of parallelization required). We fit the physical resources as a function of bit length with a function containing the terms that result from multiplying the asymptotic trends of T count and logical qubit count from table 4 in the work of H\"{a}ner et al \cite{HanerImprovedLogarithms}. The generated trend fit the data well for bit string sizes in excess of 256, with some discrepancies below this value which is likely due to the minimal degree of parallelization that was required for lower bit strings. For a code cycle time of 1 $\mu s$, base physical error rate of $10^{-3}$ and a desired run time of 0.1 day, the total physical qubits required to break the encryption as a function of bit length, n, is well described by the following equation as $0.011\times n^{4.191}$. The fit coefficients here are a function of the desired run time and code cycle time of the hardware, which dictate the required degree of parallelization. For 250 data points of bit length in the range of 100-2000, the average percentage error between the data points and the trend was $3.7\%$. The run time was maintained below 0.1 days with these hardware assumptions up to an bit length of 1730, albeit requiring $4\times 10^{11}$ physical qubits, at this point the maximum degree of parallelization was reached. Hence we show that the required number of physical qubits to break the elliptic curve encryption in a defined fixed duration scales only polynomially with the bit length of the encryption, in contrast to classical methods which require exponential time.

In figure \ref{fig:2}B we plot the required number of physical qubits to break the 256 bit elliptic curve encryption within 1 hour as a function of the base physical error rate. In section \ref{2.2} the relationship between the base physical error rate and an experimentally achieved gate fidelity is explained in more detail. We plot two trends for code cycle times equal to 1 $\mu s$ and 100 $\mu s$ which begin at the selected error rate of $10^{-5}$. The trends end at the right at a physical error rate of $\sim2.8\cdot10^{-3}$ when the AutoCCZ factory can no longer produce a state with sufficient fidelity given the error rate and number of required states. High code distance and multi tiered T gate factories would be able to continue further, but at a value of $1\%$ error rate the threshold of the surface code is reached. With a physical error rate higher than this threshold, increasing the code distance actually results in a larger logical error. Three distinct distances are calibrated according to the base physical error rate, first the final topological error is maintained below $1\%$ by adjusting the code distance associated with the data block. Next there is a level 1 code distance and a level 2 code distance associated with the AutoCCZ factory that are calibrated to ensure the final distillation error is maintained below $5\%$, as per the ancillary files of Gidney and Eker{\aa} \cite{Gidney2019FlexibleStates}. We assess a wide range of code distances for the AutoCCZ factory and choose the set that minimizes the factory volume (i.e. number of qubits $\times$ duration per cycle) while maintaining the desired error rate. With a code cycle time of 1 $\mu s$ it requires 317 million physical qubits to reach the 1 hour run time with a base error of $10^{-3}$, this is reduced down to 33 million for a base error of $10^{-4}$, i.e. a factor ~10 reduction. The relative reduction in the qubit overhead, associated with an order of magnitude improvement in the base error, is greater when the comparison is performed closer to the threshold of the code. For example from $2.8\cdot10^{-3}$ to $2.8\cdot10^{-4}$ the qubit reduction is instead a factor 30 (in contrast to the factor 10 of the previous comparison). This highlights the importance of reaching base physical error rates of $10^{-3}$ and lower.

\subsection{Finding the optimal measurement depth}\label{3.3}

Logical algorithms may be optimized for particular properties, for example, either the total T gate count, $T_{count}$, the number of measurement layers, $T_{depth}$, or logical qubit count may be minimized \cite{Amy2014Polynomial-timePartitioning,AbdessaiedTechnologyCircuits}. In the elliptic curve encryption breaking algorithm of Haner et al \cite{HanerImprovedLogarithms} logical requirements are stated for each of these three possible optimizations, where in the previous section the measurement depth minimized approach was chosen. The reaction limit (conjectured to be the fastest an algorithm can be run) is determined solely by the measurement depth and reaction time (i.e. independent of total gate count and code distance), and so the depth optimized approaches are the most suitable when room for parallelization is desired. The ratio of the measurement depth to total gate count is the inverse of the number of T gates per layer, $T_{layer}$ (when considering T gates as opposed to some other non Clifford operation). In the GoSC method of parallelization with units, all aspects of the cost depend on the number of T gates per layer, including the footprint of the unit, the time it takes to prepare a unit, and the number T gates that are effectuated within the preparation time.

\begin{figure}[t]
\centering
\includegraphics[width=17.6cm]{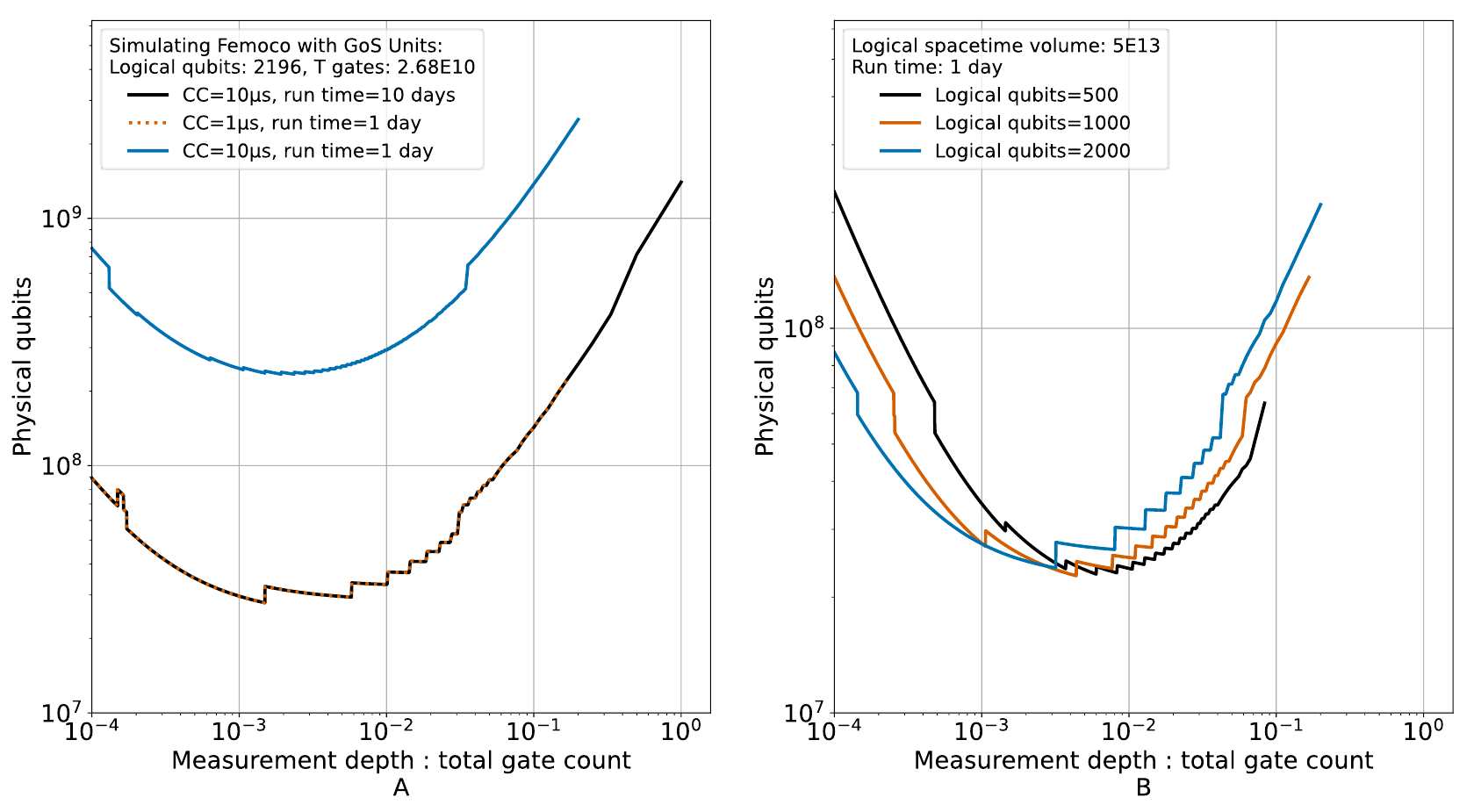}
\caption[Tof gate cost]{\label{fig:3} Investigating the efficiency of the Game of Surface Code Units approach \cite{Litinski2019gameofsurfacecodes} as a function of the ratio of the measurement depth to total gate count. The total qubit footprint is plotted for a particular desirable run time and code cycle time, with a base error of $10^{-3}$ and final output error of $6\%$. The trends start at the chosen measurement depth ratio of $10^{-4}$ and end when the reaction limit is reached before the desired run time. (A) The required number of physical qubits for a ground state energy calculation of FeMoCo with the THC Qubitization method of Lee et al \cite{Lee2020EvenHypercontraction} to have a maximum run time as stated for code cycle times of 10 $\mu s$ and 1 $\mu s$.  (B) The required number of physical qubits for an algorithm with fixed space-time volume of $5\cdot10^{13}$, where the number of logical qubits is labeled and the total T gate count is varied to maintain the stated volume. The space-time volume was chosen to be representative of the quantum advantage cases assessed in this work. }
\end{figure}

In figure \ref{fig:3} we plot the efficiency of the GoSC method as a function of this measurement depth with a fixed total gate count, and it can be seen that there is a measurement depth that leads to a minimum physical qubit footprint. This is in contrast to the AutoCCZ method where instead the measurement depth only plays a role in determining the reaction limit, and it would appear in these figures as a horizontal line. In figure \ref{fig:3}A we plot the required number of physical qubits to reach a desired run time for the logical resources required to simulate FeMoco \cite{Lee2020EvenHypercontraction}. We include three plots for different code cycle times and maximum run times. As expected in this regime prior to reaction-limited, it can be seen that it is the ratio of the code cycle time and maximum run time that determines the physical qubit requirements. We show that the optimal measurement depth ratio is not necessarily ``as small as possible", and for this situation it lies between $10^{-3}$ and $10^{-2}$. The discontinuous movements result from an increase in the required number of units as the measurement depth ratio becomes larger. A less demanding run time requirement necessitates fewer units for a given measurement depth which results in less frequent and larger relative downward movements. The two distinct discontinuous movements on the blue trend (for a code cycle time of 10 $\mu s$ and run time of 1 day) are due to the topological code distance changing as the space time volume changes. The black trend with a run time of 10 days can maintain the desired run time for larger measurement depths than the dotted orange line because the proportional difference in the desired run time (a factor 10) is greater than the proportional difference in their associated reaction times. The relationship between the code cycle time and reaction time that we assume in this work is explained in section \ref{2.3}.

In figure \ref{fig:3}B we again investigate the efficiency of the parallelization but now for an abstract algorithmic requirement with fixed space-time volume. Three trends are shown with logical qubits, $n$, corresponding to 500, 1000, and 2000, where the total T gate of the algorithm is set to maintain the fixed space time volume $(n \times T_{count})$ of $5\cdot10^{13}$. The trends end at the right when the reaction limit is reached where 2000 is the last to end because it has the lowest total T gate count, which in turn relates to a lower reaction limit for a fixed measurement depth ratio. The optimal measurement depth ratio is larger for lower logical qubit counts; for 500 the optimal is $0.6\cdot10^{-3}$, whereas for 2000 the optimal is $0.3\cdot10^{-3}$. In the following we investigate the relationship between the optimal measurement depth and logical qubit requirement of the algorithm in more detail.

\subsubsection{Optimal measurement depth and logical qubit requirement}

\begin{figure}[t!]
\centering
\includegraphics[width=17.6cm]{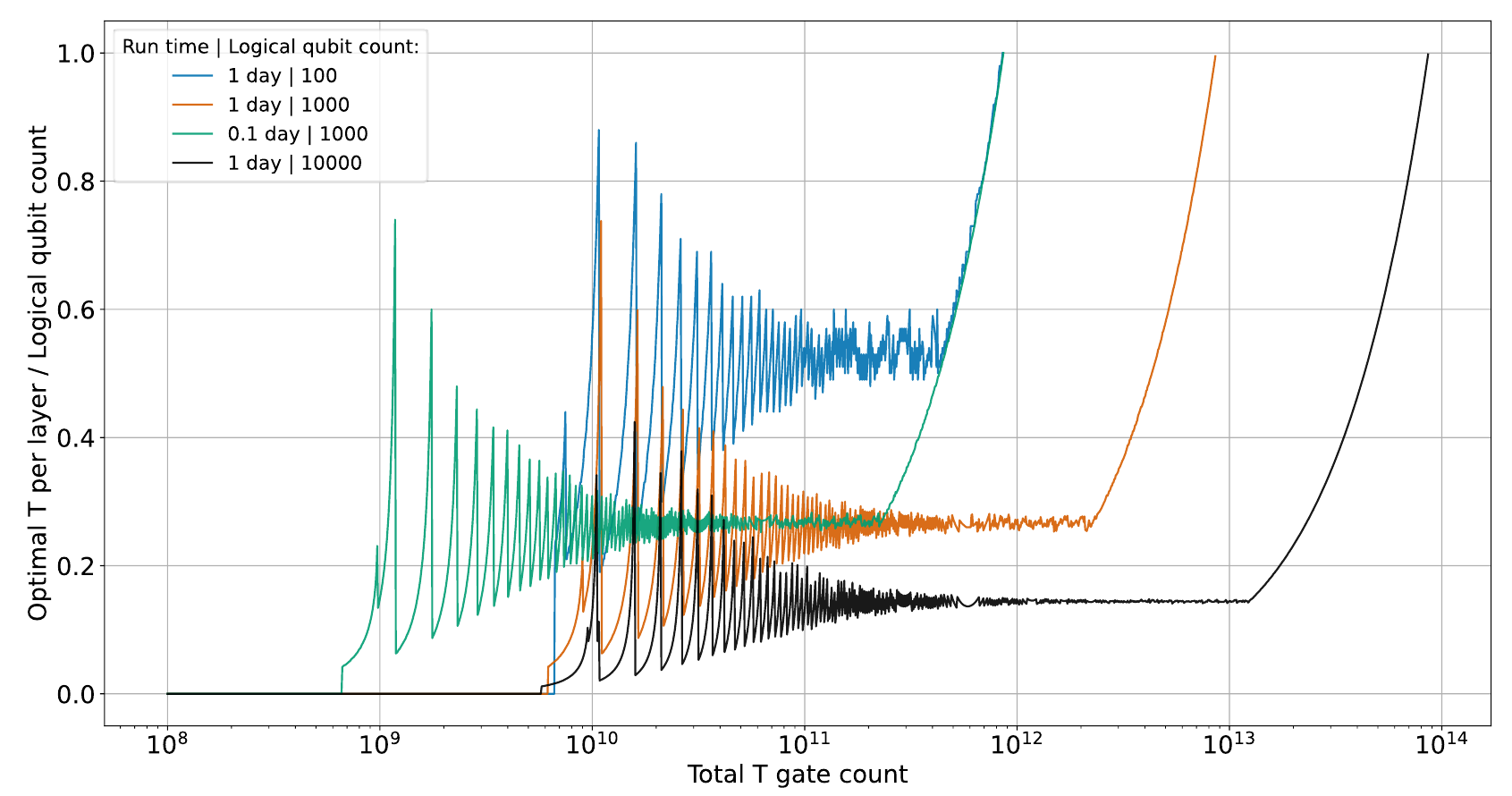}
\caption[Tof gate cost]{\label{fig:4} The optimal value of the number of T gates per measurement layer, shown as a fraction of the number of logical qubits, plotted as a function of the T gate count of the algorithm. Base physical error of $10^{-3}$ and code cycle time of 1$\mu s$. Different plots for varied desired run times and logical qubit count, 1 day and 100 qubits in blue, 1 day and 1000 qubits in orange, 1 day, 0.1 day and 1000 qubits in green, and 10,000 qubits in black. The maximum value of T gates per layer is limited to the number of logical qubits for each trend. There are 4 distinct behavioural phases for each trend, first the trends read 0 at the left while the \emph{beat} limited approach can reach the desired run time, and then units are introduced following the GoSC \cite{Litinski2019gameofsurfacecodes} method. The next phase is an oscillating pattern where the optimal T per layer ratio varies widely with a decreasing amplitude as the T gate count increases. The amplitude of the oscillation is largest at the start of parallelization because only a small number of units are initially required and an increase in T count requirement can mean the optimal number of required units increases, but the associated $T_{layer}$ for this minimum can vary widely, and the magnitude of that variation reduces as the overall number of required units increases. In the next phase the optimal T per layer ratio is relatively constant at an equilibrium value which is solely determined by the number of logical (abstract) qubits in the algorithm.  For logical qubit numbers ($N$) in excess of 100, the equilibrium value of the optimal number of T gates per layer is well described by the following equation  $T_{layer} = 1.9(5)N^{0.70(5)}$ which we estimated by taking an average of the points within the equilibrium phase and fit with linear regression. The final phase starts when the algorithmic requirement becomes too demanding for the minimum run time and so the $T_{layer}$ rises despite the loss in efficiency until the number of logical qubits is reached which here we define as the cut off point, where finally even the reaction limit (one reaction time per T layer) is not sufficient.

}
\end{figure}

We have shown that the optimal measurement depth  given a fixed total gate count for the GoSC approach is non trivial and can be a function of both the number of logical qubits, and on how demanding the desired run time is (i.e. the number of units needed). In this section we investigate the relationship between the optimal measurement depth and the number of logical (abstract) qubits of the algorithm. The average number of T gates per measurement layer, $T_{layer}$ is the inverse of the previously stated ratio of measurement depth to total gate count, i.e. $T_{layer}=T_{count}/T_{depth}$. In figure \ref{fig:4}, for a particular value of the total T gate count, $T_{count}$, we calculate the the optimal value of T gates per layer, $T_{layer}$, which is then converted into a ratio of the number of logical qubits.  We include multiple trends for different numbers of logical qubits and plot as a function of the total T gate count. This involves identifying the value of $T_{layer}$ at which the physical qubit requirement is minimized as in figure \ref{fig:3}, for each particular logical resource requirement. We choose different qubit numbers and minimum run time values to highlight the various dependencies. We can see 4 distinct behavioural phases for each trend, first at low T gate counts the minimum run time can be reached with no parallelization and so we state the optimal T per layer as 0 (whereas truly it is independent). The next phase is an oscillating pattern where the optimal T per layer ratio varies widely with a decreasing amplitude as the T gate count increases. The beginning of this phase is determined by the end of the previous one, where the no-parallelization (\emph{beat} limited) method can no longer reach the desired run time, which is determined by the ratio of the code cycle time and minimum run time. The amplitude of the oscillation is largest at the start of parallelization because only a small number of units are initially required and an increase in T count requirement can mean the number of required units increases, for example the optimal number of units may change from 3 to 4, but the associated $T_{layer}$ for this minimum can vary widely, and the magnitude of that variation reduces as the overall number of required units increases. The large variation in the optimal ratio does not imply a large variation in the required number of physical qubits. In the next phase the optimal T per layer ratio is relatively constant at an equilibrium value which is solely determined by the number of logical (abstract) qubits in the algorithm, where the greater the number of logical qubits, the fewer T gates per layer (as a percentage of the logical qubits) are required for the optimal physical qubit overhead. For logical qubit numbers ($N$) in excess of 100, the equilibrium value of the optimal number of T gates per layer is well described by the following equation  $T_{layer} = 1.9(5)N^{0.70(5)}$ which we estimated by taking an average of the points within the equilibrium phase and fit with linear regression. The final phase starts when the algorithmic requirement becomes too demanding for the minimum run time and so the $T_{layer}$ rises despite the loss in efficiency until the number of logical qubits is reached which here we define as the cut off point, where finally even the reaction limit (one reaction time per T layer) is not sufficient.

The potential degree of control over the average number of T gates per layer during algorithm construction and optimization will determine whether it is beneficial to consider the optimal value as calculated with the techniques used in figure \ref{fig:4}. There are optimization techniques that can minimize either the $T_{depth}$ and $T_{count}$ \cite{Amy2014Polynomial-timePartitioning,AbdessaiedTechnologyCircuits}, but they generally trade off with one another, where minimizing one may increase the other. Furthermore, this analysis is specific to the GoSC approach of parallelization with units, and earlier we have shown that the AutoCCZ method of parallelization produces considerably more favourable final resource estimates. As mentioned, the AutoCCZ method does not display this rich behaviour with the efficiency dependence on the measurement depth, and we believe further research is warranted to compare the underlying assumptions of these two methods of parallelization.

\section{Conclusion}

Within a particular time frame, the code cycle time and the number of achievable physical qubits may vary by orders of magnitude between hardware types. When envisaging a fault tolerant implementation, there are numerous decisions to be made based on a preference for either space or time. In this work we compare surface code strategies of parallelization that allow one to speed up the computation until the reaction limit is reached. Most of the fault tolerant resource estimation work has focused on code cycle times corresponding to superconducting architectures. A space optimized quantum advantage case study translated for hardware with slower code cycle times may lead to run times in excess of 1000 days, and so parallelization would have to be performed to reach desirable run times. In this work we have calculated the required number of physical qubits to reach a given desirable run time for two representative quantum advantage cases (chemistry and encryption) across a range of code cycle times. The feasibility of using these time optimization strategies will depend upon the number of physical qubits achievable within a device, therefore the scalability of an architecture will play an important role in determining whether a quantum advantage is achievable. We contrast two methods of parallelization to simulate the FeMo-Co cataylst, first a Game of Surface Codes approach which should be considered an upperbound, and second a more heuristic utilization of AutoCCZ factories. We find in this situation that the AutoCCZ factories produces more favourable resource estimates and the difference is mostly due to the high initial set up cost of parallelization with Game of Surface Code Units.  With a single AutoCCZ factory a superconducting device with a 1 $\mu s$ code cycle time would require 7.5 million qubits to simulate FeMo-co in $\sim$10 days, whereas a shuttling based trapped ion device with a 235 $\mu s$ code cycle time would take 2450 days. By increasing the number of factories the space-time trade is initially favourable (as opposed to linear), and the trapped ion device can reach the same 10 day run-time with 600 million qubits. In this comparison the factor difference between the physical qubit requirement is $\sim 3\times$ less than the factor difference in the code cycle time. Trapped ion architectures have generally been shown to achieve higher two qubit gate fidelities than superconducting devices, and so we have calculated that if a base physical error rate of $10^{-4}$ was achieved, the trapped ion device would instead require 60 million qubits for the 10 day run time. 

We have investigated the effect of varying the ratio of the T gate count and T depth (the average T gates per layer) and identified the optimal value for a constrained run time against general algorithmic requirements. Here we focused on the GoSC Unit method of parallelization as it was unique in displaying rich T depth dependence. 

We apply our methods to the logical resources required to break 256 elliptic curve encryption, which is used to secure public keys in the Bitcoin network. We use the logical resource requirements of the latest algorithmic developments which improve on the previous state of the art by $\sim 2$ orders of magnitude. There is a small window of time, approximately 10-60 minutes, in which the public keys are available and vulnerable after the initiation of a transaction. We quantify the number of physical qubits required to break the encryption in one hour as a function of code cycle time, and base physical error rate. It would require approximately 317 million physical qubits to break the encryption within one hour using the surface code and a code cycle time of 1 $ \mu s$, a reaction time of 10$ \mu s$, and physical gate error of $10^{-3}$. To instead break the encryption within one day, it would require \emph{only} 13 million physical qubits. If the base physical error rate was instead the more optimistic value of $10^{-4}$, 33 million physical qubits would be required to break the encryption in 1 hour.  This large physical qubit requirement implies that the Bitcoin network will be secure from quantum computing attacks for many years (potentially over a decade). Alternative error correction techniques, in particular those which benefit from a more flexible physical qubit connectivity as often found in trapped ion based quantum computers, could potentially offer considerable improvements to the requirements but the slower rate of logical operations must also be factored in. The Bitcoin network could nullify this threat by performing a soft fork onto an encryption method that is quantum secure, but there may be serious scaling concerns associated with the switch. We hope to motivate continued research into end-to-end resource estimation for alternative error correction schemes to the surface code, and to determine how best to make use of the available physical connectivity of different quantum hardware platforms.

%As expected within the parallelization regime, the depth optimized approach incurs the lowest overhead, and with a code cycle time of $1 \mu s$ and base physical error of $10^{-4}$, it would require 200 million qubits to reach the desired run time. 

\section{Acknowledgements}
This work was supported by the U.K. Engineering and Physical Sciences Research Council via the EPSRC Hub in Quantum Computing and Simulation (EP/T001062/1), the U.K.
Quantum Technology hub for Networked Quantum Information Technologies (No. EP/M013243/1), the European Commission’s Horizon-2020 Flagship on Quantum Technologies
Project No. 820314 (MicroQC), the U.S. Army Research Office under Contract No. W911NF-14-2-0106, the Office of
Naval Research under Agreement No. N62909-19-1-2116, and the University of Sussex. We thank Craig Gidney, Daniel Litinski, and Niel de Beaudrap for helpful discussions, and for their suggestions to improve the manuscript. 

\bibliography{references} 
\bibliographystyle{ieeetr}

\end{document}